\documentclass[preprint,aps,nofootinbib,showpacs,amsfonts,epsf]{revtex4-2}

\input{epsf.tex}

\newcommand{\E}{{\cal{E}}}
\newcommand{\s}{\sigma}

\renewcommand{\a}{\alpha}

\newcommand{\be}{\begin{equation}}
\newcommand{\ee}{\end{equation}}
\newcommand{\bea}{\begin{eqnarray}}
\newcommand{\eea}{\end{eqnarray}}
\newcommand{\ba}{\begin{array}}
\newcommand{\ea}{\end{array}}
\def\J#1#2#3#4{{#1} {\bf #2}, #3 (#4)}
\def\PRD{Phys. Rev. D}
\def\PR{Phys. Rev.}
\def\PRL{Phys. Rev. Lett.}

\def\PTP{Prog. Theor. Phys.}

\def\JMP{J. Math. Phys.}

\def\MZ{Math. Z.}

\def\CQG{Class. Quantum Grav.}

\def\GRG{Gen. Relativ. Grav.}

\def\MZ{Math. Zeits.}

\begin{document}
\draft
\title{On the simplest static and stationary vacuum\\ quadrupolar
metrics}

\author{I. M. Mej\'ia,$^\dagger$ V.~S.~Manko,$^\dagger$ and E.~Ruiz$^\ddagger$ }
\address{$^\dagger$Departamento de F\'\i sica, Centro de Investigaci\'on y
de Estudios Avanzados del IPN, A.P. 14-740, 07000 Ciudad de
M\'exico, Mexico\\$^\ddagger$Instituto Universitario de F\'{i}sica
Fundamental y Matem\'aticas, Universidad de Salamanca, 37008
Salamanca, Spain}

\begin{abstract}
In the present paper we argue that a special case of the Bach-Weyl
metric describing a static configuration of two Schwarzschild
black holes gives rise, after extending its parameter space to
complex values, to a very simple 2-parameter model for the
gravitational field of a static deformed mass. We compare this
model, which has no restrictions on the quadrupole parameter, with
the well-known Zipoy-Voorhees $\delta$-metric and show in
particular that the mass quadrupole moment in the latter solution
cannot take arbitrary negative values. We subsequently add an
arbitrary angular momentum to our static model and study some
properties of the resulting 3-parameter stationary solitonic
spacetime, which permits us to introduce the notion of the
Fodor-Hoenselaers-Perj\'es relativistic multipole moments.
\end{abstract}

\pacs{04.20.Jb, 04.70.Bw, 97.60.Lf}

\maketitle

\section{Introduction}

The exterior gravitational field of a static deformed mass can be
described in general relativity by various 2-parameter solutions
of Einstein's equations, and recently a comparative analysis of
some of these has been carried out in the paper \cite{FQS}. The
well-known Zipoy-Voorhees (ZV) solution \cite{Zip,Voo} (sometimes
called the $\delta$ or $\gamma$ metric) is promoted in \cite{FQS}
as the simplest quadrupole metric, advantageous over other models
in the context of physical applications, and it is worth noting
that during the last decade the ZV metric has been analyzed by
various authors. Thus, for instance, in the papers \cite{Que,BGG}
this metric was used for the analysis of the geodesic motion of
test particles in the presence of naked singularities, while the
recent work \cite{AFM} studied it in the context of quasinormal
modes of deformed compact objects; mention also that the optical
properties of the ZV metric have been analyzed in \cite{AAA}, and
the recent paper \cite{TMD} considers the harmonic oscillations of
test particles in this spacetime. However, a seemingly simple form
of the ZV metric is in reality rather deceptive as the
mass-quadrupole moment in it is a function of the varying number
of (many) Schwarzschild constituents, which in particular is
reflected in the form of the known stationary generalizations of
this static solution \cite{TSa,Yam,HKX}. Moreover, the real
situation seems to be even worse because, as will be demonstrated
in the present paper, the dimensionless quadrupole moment of the
ZV metric cannot take arbitrary negative values determining the
oblateness of the source, and it is not quite clear to which
extent this restriction persists in the stationary versions of the
$\delta$ metric too.

Motivated by the non-generic nature of the ZV solution as a
quadrupolar metric, in the present paper we will discuss another
2-parameter model of a static deformed mass, the one not mentioned
in \cite{FQS}, that arises as equatorially symmetric
specialization of the Bach-Weyl solution \cite{BWe} in which the
extension of the parameters must be additionally carried out. The
parameters of the mass and mass-quadrupole moment can be
introduced explicitly into this model instead of the original
parameter set, and the new physical parameters will have no any
restrictions on their values. We shall also consider a simple
stationary generalization of our static solution and compare it
with two known 3-parameter stationary spacetimes. This will allow
us to touch the question of the most suitable definition of the
relativistic multipole moments.

The rest of the paper is organized as follows. We start Sec.~II
with comments on the restrictions that exist in the ZV metric with
regard to the quadrupole deformations, and then construct and
analyze a simple 2-solitonic model of a static deformed mass. In
Sec.~III we consider a stationary 3-parameter generalization of
our static model and obtain its concise form in the equatorial
plane and in the extreme limit. Here we also briefly comment on
the mass-quadrupole moment in the stationary generalizations of
the ZV spacetime, and on the most appropriate definition of
multipole moments in the context of solution generating
techniques. Concluding remarks are given in Sec.~IV.

\section{The extended 2-parameter static vacuum solution}

As is well known, the static axisymmetric vacuum gravitational
fields in Einstein's general theory of relativity are described by
the Weyl line element
\be d s^2=f^{-1}[e^{2\gamma}(d\rho^2+d z^2)+\rho^2 d\varphi^2]-fd
t^2, \label{Weyl} \ee
where the functions $f$ and $\gamma$ depend on the coordinates
$(\rho,z)$ only and satisfy the differential equations
\bea &&f(f_{,\rho,\rho}+\rho^{-1}f_{,\rho}+f_{,z,z})=
f_{,\rho}^2+f_{,z}^2, \nonumber\\ &&4\gamma_{,\rho}= \rho
f^{-2}(f_{,\rho}^2-f_{,z}^2), \nonumber\\ &&2\gamma_{,z}= \rho
f^{-2}f_{,\rho}f_{,z}. \label{sfeq} \eea

The ZV solution of the system (\ref{sfeq}) has the form
\bea &&f=\left(\frac{R_++R_--2m}{R_++R_-+2m}\right)^\delta, \quad
e^{2\gamma}=\left[\frac{(R_++R_-)^2-4m^2}
{4R_+R_-}\right]^{\delta^2}, \nonumber\\
&&R_\pm=\sqrt{\rho^2+(z\pm m)^2}, \label{ZV} \eea
where $m$ and $\delta$ are two real parameters. Since the
Schwarzschild solution of mass $m$ is contained in (\ref{ZV}) as
the particular $\delta=1$ case, the constant $\delta$ may be
considered as a deformation parameter describing the deviation of
the ZV geometry from spherical symmetry.

After the redefinition $\delta=1+p$ proposed in \cite{Que} by
Quevedo, one obtains the following formulas for the total mass $M$
and mass-quadruple moment $Q$ of the ZV source \cite{Que}:
\be M=m(1+p), \quad Q=-\frac{1}{3}m^3p(1+p)(2+p), \label{MQZV} \ee
and we emphasize the importance of the `minus' sign in the
expression for $Q$.\footnote{Note that in the paper \cite{AFM} the
quadrupole moment is given with incorrect sign, which could affect
some of the results obtained in that paper.} Although at first
glance one may think that $Q$ in (\ref{MQZV}) can take arbitrary
negative values (determining the oblateness of the source)
independently of the positive values of $M$, this is not really
the case. Indeed, by inverting formulas (\ref{MQZV}) and
redefining $Q=qM^3$, we get
\be m=M\sqrt{1+3q}, \quad p=-1+\frac{1}{\sqrt{1+3q}}, \label{mp}
\ee
whence it follows, taking into account the reality of the constant
$p$, that the negative values of the dimensionless quadrupole
moment $q$ are restricted by the inequality
\be -\frac{1}{3}<q<0, \label{qZV} \ee
the corresponding values of $p$ varying from $0$ to $+\infty$.
This obviously invalidates the ZV metric as a generic 2-parameter
model for the exterior field of a static deformed mass.

It appears that the well-known Bach-Weyl (BW) solution \cite{BWe}
for two nonequal Schwarzschild masses permits us to elaborate a
more attractive model for a static deformed mass in which the
mass-quadrupole moment would not have already any restrictions on
its negative values. A starting point in the construction of our
model is a special case of the BW spacetime in which the
separation parameter is set equal to zero, so that the resulting
configuration of two overlapping sources becomes symmetric with
respect to the equatorial plane, the corresponding metric
functions $f$ and $\gamma$ having the form
\bea f&=&\frac{(R_++R_--2m_1)(r_++r_--2m_2)}
{(R_++R_-+2m_1)(r_++r_-+2m_2)}, \nonumber\\
e^{2\gamma}&=&\frac{(m_1+m_2)^2[(R_++R_-)^2-4m_1^2]
[(r_++r_-)^2-4m_2^2]} {16(m_1-m_2)^2R_+R_-r_+r_-} \nonumber\\
&&\times\left[\frac{m_2(R_++R_-)-m_1(r_++r_-)}
{m_2(R_++R_-)+m_1(r_++r_-)}\right]^2, \nonumber\\
R_\pm&=&\sqrt{\rho^2+(z\pm m_1)^2}, \quad r_\pm=\sqrt{\rho^2+(z\pm
m_2)^2}. \label{BW} \eea
The two arbitrary real parameters of this solution are $m_1$ and
$m_2$, and by analogy with the general case they can be
interpreted as individual masses of the Schwarzschild
constituents. On the upper part of the symmetry axis ($\rho=0$,
$z>{\rm max}\{m_1,m_2\}$) the function $\gamma$ vanishes, while
the function $f$ takes the form
\be f(\rho=0,z)=\frac{(z-m_1)(z-m_2)}{(z+m_1)(z+m_2)}, \label{fax}
\ee
whence we readily get, via the procedure \cite{FHP} of calculating
the Geroch-Hansen (GH) multipole moments \cite{Ger,Han}, the
expressions for the total mass and mass-quadrupole moment:
\be M=m_1+m_2, \quad Q=-m_1m_2(m_1+m_2). \label{MQBW} \ee
Now, inverting the above formulas and introducing the
dimensionless quadrupole moment $q=Q/M^3$, we get
\be m_1=\frac{M}{2}(1+d), \quad m_2=\frac{M}{2}(1-d), \quad
d=\sqrt{1+4q}, \label{mq} \ee
and hence, accounting for the reality of $m_1$ and $m_2$, one
might think that the allowed negative values of $q$ lie in the
interval $(-{\textstyle\frac{1}{4}},0)$, thus being even more
restrictive than in the case of the ZV solution. However, it is
easy to see that for all $q<-{\textstyle\frac{1}{4}}$ the
parameters $m_1$ and $m_2$ in (\ref{mq}) become complex conjugate
quantities, $m_2=\bar m_1$, and these preserve the reality of the
axis expression (\ref{fax}), so that the reality of the metric
functions $f$ and $\gamma$ in (\ref{BW}) is also preserved.
Therefore, after changing in (\ref{BW}) the parameters $(m_1,m_2)$
to $(M,q)$ by means of (\ref{mq}), we finally arrive at the
extended version of our 2-parameter model satisfying the system
(\ref{sfeq}), in which the quadrupole moment $q$ can take
arbitrary real values:
\bea f&=&\frac{[R_++R_--M(1+d)][r_++r_--M(1-d)]}
{[R_++R_-+M(1+d)][r_++r_-+M(1-d)]}, \nonumber\\
e^{2\gamma}&=&\frac{[(R_++R_-)^2-M^2(1+d)^2]
[(r_++r_-)^2-M^2(1-d)^2]} {16(1+4q)R_+R_-r_+r_-} \nonumber\\
&&\times\left[\frac{(1-d)(R_++R_-)-(1+d)(r_++r_-)}
{(1-d)(R_++R_-)+(1+d)(r_++r_-)}\right]^2, \nonumber\\
R_\pm&=&\sqrt{\rho^2+[z\pm M(1+d)/2]^2}, \quad
r_\pm=\sqrt{\rho^2+[z\pm M(1-d)/2]^2}. \label{BWq} \eea
Note that in terms of the new parameters $M$ and $q$ the axis
expression (\ref{fax}) takes the extended form
\be f(\rho=0,z)=\frac{z^2-Mz-M^2q}{z^2+Mz-M^2q}, \label{faxq} \ee
thus being arguably the most concise and elegant axis expression
of the function $f$ containing explicitly the mass monopole and
quadrupole relativistic moments as arbitrary parameters. To the
best of our knowledge, the extension of the parameters $m_1$ and
$m_2$ to the complex values in the BW solution has never been
attempted earlier in the literature, most probably because such an
extension involving `imaginary' masses might be erroneously taken
for unphysical by the researchers.

It is clear that the extended solution (\ref{BWq}) contains the
entire non-extended BW solution (\ref{BW}) as a particular
subextreme case defined by $q>-\textstyle{\frac{1}{4}}$, and in
the limit $q=-\textstyle{\frac{1}{4}}$ it reduces to the ZV
$\delta=2$ spacetime. The hyperextreme part of our solution
corresponds to $q<-\textstyle{\frac{1}{4}}$, in which case the
functions $R_\pm$ and $r_\pm$ take complex values, and it is worth
noting in this respet that the condition for fixing uniquely the
branch of the square roots $r_i=\sqrt{\rho^2+(z-\a_i)^2}$ is ${\rm
Re}(r_i)>0$, independently of whether $\a_i$ are real or imaginary
quantities.

\section{The extended 3-parameter stationary vacuum solution}

We now turn to discussing stationary generalizations of the
2-parameter solutions considered in the previous section.
Concerning the ZV metric, the main question to examine is whether
the introduction of the angular momentum is able to affect somehow
the admissible negative values of its mass-quadrupole moment; on
the other hand, the static model (\ref{BWq}) is likely to be given
the simplest possible generalization to the stationary case.

Undoubtedly, the most renowned stationary generalization of the ZV
metric is the Tomimatsu-Sato (TS) family \cite{TSa,Yam} of
solutions for spinning masses which was originally constructed for
integral $\delta$ only, and later extended to arbitrary real
$\delta$ too \cite{Yam2,Hor}. The complexity of the TS solutions
grows rapidly with growing $\delta$ \cite{Yam}, which makes them
hardly recommended for the use as a simple model describing the
field of a spinning mass with arbitrary quadrupole deformation. At
the same time, it is not difficult to demonstrate that the
mass-quadrupole moment in this important family of stationary
spacetimes defines larger oblateness of the source than in the
static ZV solution. Indeed, the total mass $M$, the quadrupole
moment $Q$ and the total angular momentum $J$ of the TS solutions
are given by the formulas \cite{SKM,Man}
\be M=\frac{\delta\s}{p_0}, \quad Q=-M^3\left(
\frac{\delta^2-1}{3\delta^2}p_0^2+q_0^2\right), \quad J=M^2q_0,
\label{QTS} \ee
where $\s$ is an arbitrary positive constant, while the real
parameters $p_0$ and $q_0$ are subject to the constraint
$p_0^2+q_0^2=1$. Solving the system (\ref{QTS}) for $\delta$, $\s$
and $q_0$ and taking into account that $q_0$ represents the
dimensionless angular momentum, we obtain, after introducing
$q=Q/M^3$ and redefining $\delta=1+p$, a stationary generalization
of formulas (\ref{mp}):
\be \s=M\sqrt{1+3q+2q_0^2}, \quad
p=-1+\frac{\sqrt{1-q_0^2}}{\sqrt{1+3q+2q_0^2}}, \label{sp} \ee
whence it follows at once that the negative values of the
dimensionless quadrupole moment $q$ are determined by the
inequalities
\be -\frac{1}{3}(1+2q_0^2)<q<0, \quad 0\le q_0^2<1, \label{qTS}
\ee
and hence the angular momentum in the TS solutions indeed enlarges
the oblateness of the massive sources.

Another well-known stationary generalization of the ZV metric is
the Hoenselaers-Kinnersley-Xanthopoulos (HKX) solution
\cite{HKX,Yam3}, some variations of which have been analyzed by
different authors in application to various problems involving a
spinning deformed mass \cite{Que2,TQu,FSo,AFM2}. The simplest
asymptotically flat 3-parameter HKX solution possessing equatorial
symmetry is defined by the Ernst complex potential $\E$ \cite{Ern}
of the form ($p=\delta-1$)
\bea \E&=&\left(\frac{x-1}{x+1}\right)^p\frac{A_-}{A_+}, \nonumber\\
A_\mp&=&(x\mp1)(x^2-1)^{2p}-\a^2(x\pm1)(x^2-y^2)^{2p} \nonumber\\
&&-i\a(x^2-1)^p[(y\pm1)(x+y)^{2p}+(y\mp1)(x-y)^{2p}], \label{EHKX}
\eea
where the real constant $\a$ is the rotation parameter, and the
spheroidal coordinates $x$ and $y$ are related to the coordinates
$\rho$ and $z$ by the formulas
\be x=\frac{1}{2\s}(r_++ r_-), \quad y=\frac{1}{2\s}(r_+-r_-),
\quad r_\pm=\sqrt{\rho^2+(z\pm\s)^2}, \label{xy} \ee
$\s$ being a positive real constant. The ZV solution is contained
in (\ref{EHKX}) as the particular $\a=0$ case.

Although the HKX version of a stationary $\delta$-metric is more
compact and simple than the respective TS version with nonintegral
$\delta$, the expressions of the physical quantities $M$, $J$ and
$Q$ defined by (\ref{EHKX}) turn out to be more complicated than
formulas (\ref{QTS}) of the generalized TS family. This can be
seen with the aid of the form of the potential (\ref{EHKX}) on the
upper part of the symmetry axis ($\rho=0$, $z>\s$), namely,
\be \E(\rho=0,z)=\frac{(z-\s)^p[z-\s-\a^2(z+\s)]-2i\s\a(z+\s)^p}
{(z+\s)^p[z+\s-\a^2(z-\s)]-2i\s\a(z-\s)^p}, \label{HKXaxis} \ee
whence the desired multipole moments of the HKX spacetime can be
obtained by means of the procedure \cite{FHP}, finally yielding
\bea M&=&\s\left(p+\frac{1+\a^2}{1-\a^2}\right), \nonumber\\
J&=&\frac{2\s^2\a[1+\a^2+2p(1-\a^2)]}{(1-\a^2)^2}, \nonumber\\
Q&=&-\frac{\s^3}{3(1-\a^2)^3}\{12\a^2(1+\a^2)
+p(1-\a^2)[(1+p)(2+p) \nonumber\\
&&+2\a^2(16-p^2)+\a^4(1-p)(2-p)]\}, \label{QHKX} \eea
and it seems impossible to resolve the algebraic system
(\ref{QHKX}) analytically for $\s$, $p$ and $\a$. Nonetheless, it
is still possible to demonstrate numerically that, thanks to
rotation, the oblateness of the HKX stationary source can be
larger than that of the ZV static source. For this purpose, one
has first to pass in (\ref{QHKX}) to the dimensionless angular
momentum and quadrupole moment, $j=J/M^2$ and $q=Q/M^3$
respectively, and then assign particular values to $M$, $j$ and
$q$ in order to finally get the corresponding meaningful values of
$\s$, $p$ and $\a$ through the resolution of the system
(\ref{QHKX}) numerically. Thus, for the particular choice $M=1$,
$j=0.65$, $q=-0.5$ we obtain $\s\approx0.562$, $p\approx0.406$,
$\a\approx0.397$ (numerical values are given up to three decimal
places); by further leaving $M$ and $q$ unchanged and varying only
$j$, we find for $j=0.68$ the numerical solution $\s\approx0.645$,
$p\approx0.172$, $\a\approx0.399$, while the value $j=0.7$ gives
$\s\approx0.692$, $p\approx0.043$, $\a\approx0.409$. Since
$q=-0.5$ in the above numerical solutions is greater in absolute
value than the absolute value of the minimal dimensionless
quadrupole moment $q$ of the ZV metric, we have shown, on the one
hand, that oblateness of the HKX spinning source can be larger
compared with that of the ZV static source. On the other hand, the
dependence $p(j)$ in the above examples, when a smaller $p$
corresponds to a larger $j$, simply means that a static ZV source
with a smaller intrinsic oblateness needs a larger angular
momentum to be deformed beyond the limiting value $-1/3$ than the
static source with a larger intrinsic oblateness, which looks
quite natural.

We would like to emphasize that both the TS and the HKX families
of solutions are physically meaningful spacetimes which over the
years have been widely discussed in the literature as legitimate
examples representing the exterior field of a spinning mass. At
the same time, it is also clear that these solutions can hardly be
advocated as the simplest generic models for the exterior geometry
around compact spinning objects, with advantages over other known
exact solutions. Actually, in what follows we are going to point
out a special member of the extended 2-soliton stationary solution
which generalizes in a very simple way the static solution
(\ref{BWq}) from the previous section and contains explicitly the
multipoles $M$, $J$ and $Q$ as three arbitrary real parameters.

We note that in the paper \cite{MRu} a physical representation of
the general 4-parameter metric for the exterior field of a neutron
star was obtained in terms of multipole moments. So, taking into
account that the static solution (\ref{BWq}) is a 2-parameter
specialization of that metric, it would be logic to search for its
simplest stationary generalization within the same generic metric
too. A thorough analysis of the axis expression of the Ernst
potential defining the general 4-parameter solution has eventually
led us to the particular 3-parameter spacetime with the following
remarkably simple axis data:
\be \E(\rho=0,z)\equiv
e(z)=\frac{z^2-Mz-M^2q-iM^2j}{z^2+Mz-M^2q+iM^2j}, \label{E3axis}
\ee
where the parameters $q$ and $j$, as before, are the dimensionless
mass-quadrupole moment and dimensionless angular momentum,
respectively. The potential $\E$ of the new 3-parameter extended
solution in the entire space has the form
\bea \E&=&(A-B)/(A+B), \nonumber\\
A&=&(\s_++\s_-)^2(R_+-R_-)(r_+-r_-)
-4\s_+\s_-(R_++r_-)(R_-+r_+), \nonumber\\
B&=&Md[\s_-(R_+-R_-)+\s_+(r_+-r_-)], \nonumber\\
R_\pm&=&\frac{\pm\s_++ij}{d+1} \sqrt{\rho^2+\left(z\pm
M\s_+\right)^2}, \quad r_\pm=\frac{\pm\s_-+ij}{d-1}
\sqrt{\rho^2+\left(z\pm M\s_-\right)^2},
\nonumber\\
\s_\pm&=&\sqrt{q+(1\pm d)/2}, \quad d=\sqrt{1+4(q+j^2)},
\label{E3p} \eea
and it satisfies the Ernst equation \cite{Ern}
\be (\E+\bar\E)(\E_{,\rho,\rho}+\rho^{-1}\E_{,\rho}+\E_{,z,z})=
2(\E_{,\rho}^2+\E_{,z}^2). \label{Eeq} \ee

The corresponding full metric is given by the line element
\be d s^2=f^{-1}[e^{2\gamma}(d\rho^2+d z^2)+\rho^2 d\varphi^2]-f(d
t-\omega d\varphi)^2, \label{Papa} \ee
with the metric functions $f$, $\gamma$ and $\omega$ defined by
the expressions
\bea f&=&\frac{A\bar A-B\bar B}{(A+B)(\bar A+\bar B)}, \quad
e^{2\gamma}=\frac{A\bar A-B\bar B}{K_0\bar K_0R_+R_-r_+r_-}, \quad
\omega=-\frac{{\rm 2Im}[G(\bar
A+\bar B)]}{A\bar A-B\bar B}, \nonumber\\
G&=&Md[\s_-(R_+-R_-)(r_++r_--z)-\s_+(r_+-r_-)(R_++R_-+z) \nonumber\\
&&+M\s_+\s_-(R_++R_-+r_++r_-)], \nonumber\\
K_0&=&4d^2\s_+\s_-/(q+j^2), \label{m3p} \eea
which have been worked out with the aid of the general formulas of
the paper \cite{MRu}. Note that the Kerr metric \cite{Ker} is
contained in the above formulas as the $q=-j^2$ particular case.
Note also that the singularities of the solution are defined by
the equation $A+B=0$, and they lie on the stationary limit surface
$f=0$, which means in particular that the singularities are not
expected to show up when the solution is used as the exterior part
in a physically meaningful global model of a compact object.

The solution (\ref{E3p}) has a remarkable multipole structure. Its
first four mass and angular momentum relativistic moments,
obtainable from (\ref{E3axis}), have the form
\bea &&M_0=M, \quad M_1=0, \quad M_2=M^3q, \quad M_3=0,
\nonumber\\ &&J_0=0, \quad J_1=M^2j, \quad J_2=0, \quad J_3=M^4qj,
\label{mm4} \eea
and it can be shown that the complex coefficients $m_n$ in the
expansion of the function
\be \xi(z)\equiv\frac{1-e(z)}{1+e(z)}=
\sum\limits_{n=0}^{\infty}m_nz^{-n-1}, \label{xiz} \ee
when $z\to\infty$, which play a key role in the procedure
\cite{FHP}, are defined in the case of the solution (\ref{E3p}) by
the following very concise generic formulas:
\be m_{2k}=M^{2k+1}q^k, \quad m_{2k+1}=iM^{2k+2}q^kj, \quad
k=0,1,2,\ldots \label{mn3p} \ee

It is well known \cite{FHP} that only the first four quantities
$m_n$ coincide with the GH complex multipoles $P_n=M_n+iJ_n$, the
Kerr solution being an exclusive stationary vacuum spacetime for
which $P_n=m_n$ for all $n$, so that we easily get from
(\ref{mn3p}), after setting $q=-j^2$, the simple formula defining
the multipole moments of the Kerr spacetime \cite{Han}:
\be P_{n}=M(iMj)^{n}, \quad n=0,1,2,\ldots \label{momKer} \ee
When $q\ne-j^2$, the calculation of the GH multipole moments of
the solution (\ref{E3p}) in the general case requires finding
corrections to the coefficients $m_n$, $n\ge4$, which would
involve combinations of the lower multipoles. Since these
additional terms must inevitably spoil the exceptional multipole
structure (\ref{mn3p}) determined by the progressive powers of
$q$, we find it necessary to briefly reexamine the issue of
multipole moments in the context of the modern solution generating
techniques.

We first note that the GH multipole moments were defined as values
of certain tensorial quantities ``at infinity'', and their
practical calculation actually represents a complicated task. That
is why Fodor {\it et al.} in their paper \cite{FHP} developed for
the stationary axisymmetric asymptotically flat spacetimes an
algorithm, based on expansion of the axis expression of the Ernst
complex potential $\xi$, which considerably simplifies the
computation of the GH multipoles $P_n$. The authors of \cite{FHP}
magnanimously interpreted the discrepancies between the quantities
$m_n$ and $P_n$ in favor of the latter, and found the explicit
form of the first eleven $P_n$ in terms of $m_n$; however, they
failed to discern in $m_n$ an independent definition of
relativistic multipole moments which has various advantages over
the GH one. From now on we will refer to the complex quantities
$m_n$ as the Fodor-Hoenselaers-Perj\'es (FHP) multipole moments,
the real part of which determines the mass multipoles and the
imaginary part -- the rotational multipoles. It should be noted
that the axis value of the Ernst complex potential is the key
ingredient in the modern solution generating techniques, and its
FHP multipoles $m_n$ determine it uniquely; the subsequent
construction of the solution in the whole space can be carried out
for instance with the aid of Sibgatullin's integral method
\cite{Sib}. At the same time, the construction of an exact
solution with a specified number of GH multipole moments would
require identification of the corresponding infinite set of the
FHP moments $m_n$ as a preliminary step for finding the respective
axis data, which generically does not look realizable in
principle. Our solution (\ref{E3p}) illustrates well the above
said: we had no problems in finding the general formula for the
FHP multipoles (\ref{mn3p}), while we see no way of getting the
corresponding general expression for $P_n$, even though the axis
data (\ref{E3axis}) contains only three parameters.

The adoption of the FHP multipole moments instead of the GH ones
makes it possible to rectify the physical interpretation of some
known exact solutions. For example, the limit $q=0$ in the
solution (\ref{E3p}) leads to the 2-parameter $M$-$j$ spacetime
possessing only two nonzero FHP moments $m_0$ and $m_1$, and
therefore describing the exterior gravitational field of a rigidly
rotating sphere. Interestingly, this particular case of the
extended 2-soliton solution \cite{MMR} was briefly analyzed in the
paper \cite{HMR} and found physically deficient due to a specific
behavior of its higher GH multipole moments. Since the Ernst
potential of the $M$-$j$ solution was given in \cite{HMR} with
errors, it will be worth noting that the case $q=0$ in formulas
(\ref{E3p}) is determined by $d$ and $\s_\pm$ of the form
\be d=\sqrt{1+4j^2}, \quad \s_\pm=\sqrt{(1\pm d)/2}
=\frac{1}{2}\bigl(\sqrt{1+2ij}\pm\sqrt{1-2ij}\bigr). \label{ds}
\ee
It is clear that within the framework of the GH definition of
multipole moments the correct physical interpretation of this
special case would be impossible.

The 3-parameter solution (\ref{E3p}) has a simple form suggesting
the suitability of the solution for astrophysical applications and
gravitational experiment. Since a considerable part of such
applications is restricted to the analysis of different physical
processes and effects in the equatorial plane, it is desirable to
have the representation of the metric functions (\ref{m3p}) of our
solution in the limit $z=0$. It is not difficult to show, using
the results of \cite{MRu}, that the $M$-$q$-$j$ metric in the
equatorial plane takes the following concise form:
\bea f&=&\frac{{\mathcal A}-{\mathcal B}}{{\mathcal A}+{\mathcal
B}}, \quad e^{2\gamma}=\frac{{\mathcal A}^2-{\mathcal
B}^2}{({r}_++{r}_-)^4{r}_+^2{r}_-^2}, \quad \frac{\omega}{M}=
-\frac{2j{\mathcal B}}{{\mathcal A}-{\mathcal B}}, \nonumber\\
{\mathcal A}&=&({r}_++{r}_-)^2
{r}_+{r}_-+q+j^2, \nonumber\\
{\mathcal B}&=&({r}_++{r}_-) ({r}_+{r}_-+r+q), \nonumber\\
r_\pm&=&\sqrt{r+q+\frac{1}{2}\left[1\pm\sqrt{1+4(q+j^2)}\right]},
\quad r\equiv\rho^2/M^2, \label{mfz0} \eea
and, remarkably, all three metric functions in (\ref{mfz0}) are
determined exclusively by the $z=0$ value of the Ernst potential.

We also note that the extreme limit in the solution (\ref{E3p})
occurs when $d=0$ $\Leftrightarrow$
$q=-\textstyle{\frac{1}{4}}-j^2$, and the form of the solution
then can be worked out with the aid of the general formulas of
Ref.~\cite{MRu2}. In this case it is convenient to introduce the
spheroidal coordinates $x$ and $y$ by the formulas
\bea x&=&\frac{1}{2M\s}(r_++ r_-), \quad
y=\frac{1}{2M\s}(r_+-r_-), \nonumber\\
r_\pm&=&\sqrt{\rho^2+(z\pm M\s)^2}, \quad
\s=\sqrt{\frac{1}{4}-j^2}, \label{xy} \eea
in terms of which the potentials $\E$ can be written as
\bea \E&=&(A-B)/(A+B), \nonumber\\
A&=&\lambda^2+\frac{1}{2}\lambda +ij\s xy(1-y^2),
\nonumber\\
B&=&(\s x+ijy)\lambda+\frac{1}{2}ijy(1-y^2), \label{Eext} \eea
while for the metric coefficients $f$, $\gamma$ and $\omega$ we
readily get the expressions
\bea f&=&\frac{N}{D}, \quad e^{2\gamma}=\frac{N}{\s^8(x^2-y^2)^4},
\quad \omega=\frac{M(y^2-1)W}{N}, \nonumber\\
N&=&\lambda^4-\s^2(x^2-1)(1-y^2)\nu^2, \nonumber\\
D&=&N+\lambda^2\kappa+(1-y^2)\nu\chi, \nonumber\\
W&=&\s^2(x^2-1)\nu\kappa+\lambda^2\chi, \nonumber\\
\lambda&=&\s^2(x^2-1)-j^2(1-y^2), \nonumber\\
\nu&=&jy^2, \nonumber\\
\kappa&=&2\s^2(\s x+1)(x^2-y^2)+\frac{1}{2}[\s
x(y^2+1)+y^2], \nonumber\\
\chi&=&2j\s^2(\s x+1)(x^2-y^2)+\frac{1}{2}j\s x(1-y^2).
\label{mfext} \eea

In view of a not quite accurate statement made in the paper
\cite{MRu2} concerning the relation of the extreme vacuum
potential (24) of \cite{MRu2} to the well-known Kinnersley-Chitre
(KC) 5-parameter solution \cite{KCh}, we find it instructive to
reexamine this issue in more detail. As we have been able to find
out recently, the desired relation can be only established for
nonzero values of the KC parameter $\beta$, which may look strange
recalling that this parameter is responsible for counterrotation
(see, e.g., \cite{Yam4}), while the solution (24) of \cite{MRu2},
as well as the solution (\ref{Eext}) of this paper, is
equatorially symmetric. However, as we have discovered to our big
surprise, under certain choices of other parameters in the KC
solution, $\beta$ can describe the corotating case too. Thus, by
setting $\a=Q=0$, $P=1$, $\beta=-p_0q_0$ in the KC solution for
which we use notations of Ref.~\cite{MRu3}, we get the following
axis value of the corresponding Ernst potential:
\be \E(\rho=0,z)=\frac{e_-}{e_+}, \quad e_\mp=z^2\mp
\frac{2\s}{p_0}z +(1+q_0^2)\frac{\s^2}{p_0^2}\pm
2iq_0\frac{\s^2}{p_0^2}, \label{KCaxis} \ee
where $p_0$ and $q_0$ satisfy the relation $p_0^2+q_0^2=1$, and
$\s$ is a real constant. Then, taking into account that the total
mass $M$ and angular momentum $J$ of the KC solution in the
particular case (\ref{KCaxis}) are defined by the formulas
$M=2\s/p_0$ and $J=-M^2q_0/2$, we can introduce $j=-q_0/2$ and
thus arrive at the axis expression of the solution (\ref{Eext}) in
which the values of the dimensionless angular momentum $j$ are
restricted by the inequality $|j|<1/2$ since $|q_0|\le1$.
Therefore, the extreme solution (\ref{Eext}) of this section, and
the solution (24) of \cite{MRu2} represent the analytically
extended versions of the KC subcase (\ref{KCaxis}). Mention, that
the values $|j|>1/2$ can be also covered by the KC solution, but
only after a complex continuation of the parameters $p\to ip$,
$\s\to i\s$, $q_0^2-p_0^2=1$, so that $|q_0|\ge1$.

We emphasize that the extreme metric (\ref{Eext})-(\ref{mfext})
describes an oblate object.

\section{Conclusions}

In the present paper we have shown that the ZV metric, regarded by
various authors as the simplest model for a static deformed mass,
in reality has restrictions on the negative values of the
dimensionless quadrupole moment and as such can hardly be
considered superior to other known 2-parameter solutions for a
non-spherical mass \cite{ERo,GMa,Man2,HMa}. At the same time, in
the context of the ``simplest model'' we have proposed a static
2-parameter metric, obtainable as analytic extension of the BW
spacetime, in which the two arbitrary parameters are explicitly
the mass monopole and quadrupole moments. This metric was
subsequently generalized to the stationary case, to include the
angular momentum parameter. The analysis of the multipole
structure of our 3-parameter spacetime has led us to the
conclusion that the GH multipole moments distort generically the
physical interpretation of exact solutions and therefore have to
be substituted for what we have called the FHP multipoles, the
latter being better adjusted to the modern solution generating
methods and to the intrinsic structure of stationary axisymmetric
spacetimes. In this respect we would like to observe that Hansen
himself \cite{Han} pointed out the ambiguity in the definition of
multipole moments, so that there should be no surprise when a more
precise definition eventually replaces the old one.

Apparently, the use of the FHP multipole moments instead of the GH
ones will be able to considerably simplify the multipole analysis
of the stationary axially symmetric solutions, facilitating in
particular comparison of the analytical and numerical models of
astrophysical interest. It is worth noting in this regard that the
FHP mass-hexadecapole moment of the 4-parameter solution for the
exterior geometry of a neutron star \cite{MRu} will be strictly
quartic in angular momentum, thus lending full support to the Yagi
{\it et al.} no-hair conjecture for neutron stars \cite{YKP}.

\section*{Acknowledgments}

We thank the referee for valuable remarks. This work was partially
supported by CONACYT of Mexico, by Project PGC2018-096038-B-100
from Ministerio de Ciencia, Innovaci\'on y Universidades of Spain,
and by Project SA083P17 from Junta de Castilla y Le\'on of Spain.

\end{document}